# Entropic evidence for a Pomeranchuk effect in magic angle graphene

Asaf Rozen[1†], Jeong Min Park[2†], Uri Zondiner[1†], Yuan Cao[2†], Daniel Rodan-Legrain[2], Takashi Taniguchi[3], Kenji Watanabe[3], Yuval Oreg[1], Ady Stern[1], Erez Berg[1*], Pablo Jarillo-Herrero [2*] and Shahal Ilani[1*]

[1] *Department of Condensed Matter Physics, Weizmann Institute of Science, Rehovot 76100, Israel.*
[2] *Department of Physics, Massachusetts Institute of Technology, Cambridge, Massachusetts 02139, USA.*
[3] *National Institute for Materials Science, 1-1 Namiki, Tsukuba, 305-0044 Japan.*
[†] These authors contributed equally to the work.
[*] Correspondence to: erez.berg@weizmann.ac.il, pjarillo@mit.edu, shahal.ilani@weizmann.ac.il

**In the 1950's, Pomeranchuk[1] predicted that, counterintuitively, liquid ³He may solidify upon heating, due to a high excess spin entropy in the solid phase. Here, using both local and global electronic entropy and compressibility measurements, we show that an analogous effect occurs in magic angle twisted bilayer graphene. Near a filling of one electron per moiré unit cell, we observe a dramatic increase in the electronic entropy to about $1k_B$ per unit cell. This large excess entropy is quenched by an in-plane magnetic field, pointing to its magnetic origin. A sharp drop in the compressibility as a function of the electron density, associated with a reset of the Fermi level back to the vicinity of the Dirac point, marks a clear boundary between two phases. We map this jump as a function of electron density, temperature, and magnetic field. This reveals a phase diagram that is consistent with a Pomeranchuk-like temperature- and field-driven transition from a low-entropy electronic liquid to a high-entropy correlated state with nearly-free magnetic moments. The correlated state features an unusual combination of seemingly contradictory properties, some associated with itinerant electrons, such as the absence of a thermodynamic gap, metallicity, and a Dirac-like compressibility, and others associated with localized moments, such as a large entropy and its disappearance with magnetic field. Moreover, the energy scales characterizing these two sets of properties are very different: whereas the compressibility jump onsets at $T \sim 30\text{K}$, the bandwidth of magnetic excitations is $\sim 3\text{K}$ or smaller. The hybrid nature of the new correlated state and the large separation of energy scales have key implications for the physics of correlated states in twisted bilayer graphene.**

Systems of strongly interacting fermions exhibit a competition between localization, minimizing the potential energy, and itinerancy, minimizing the kinetic energy. The advent of two-dimensional moiré systems, such as magic angle twisted bilayer graphene[2–6] (MATBG), opens a new route to study this physics by controlling the ratio between the electronic interactions and bandwidth in a highly tunable artificial lattice. In systems where this ratio is large, such as transition metal dichalcogenides hetero-bilayers, electrons tend to localize to the lattice sites, forming Mott insulators[7,8]. In the other extreme, where the electronic bandwidth is large, as in bilayer graphene with a large twist angle, a Fermi liquid state is formed in which electrons are itinerant. MATBG provides a fascinating example of a system at the boundary between these two extremes. This system shows a host of electronic phases, including correlated insulators[3,9,10], Chern insulators[11–13], superconductors[4,9,10], and ferromagnets[14,15]. Scanning tunneling spectroscopy[16–19] and electronic compressibility measurements[20,21] indicate that in this system Coulomb interactions and kinetic energies are indeed comparable. In this regime, there is an inherent tension between localized and itinerant descriptions of the physics. Moreover, the growing understanding that the nearly-flat bands in MATBG have a topological character[22–24] implies that a simple "atomic" description, in which electrons are localized to individual moiré lattice sites, may not be appropriate. Instead, a picture analogous to that of quantum Hall ferromagnetism has been proposed as an alternative starting point[25–27]. Understanding this interplay between itinerancy and localization, and the new physics that emerges from it, remains a major challenge.

In this work we find that, surprisingly, the correlated state in MATBG above a filling of one electron per moiré site has a hybrid nature, with some of its properties resembling those of an itinerant system, and others which are usually associated with localized electrons. Measurements of the electronic entropy at temperatures as low as a few Kelvin reveal that this state has an unusually large excess entropy, which is rapidly suppressed by a moderate in-plane magnetic field. This suggests that even at such low temperatures, there are strongly fluctuating magnetic moments in the system, a behavior that is typically associated with local moments. On the other hand, our measurements find that this state is metallic and has no thermodynamic gap nearby, which is naturally understood within an itinerant picture.

The presence of fluctuating moments at temperatures that are much smaller than the electronic bandwidth indicates the existence of a new, anomalously small energy scale associated with the bandwidth of magnetic excitations, which is an order of magnitude smaller than the energy scale where a jump appears in the compressibility[21,28]. This jump marks the boundary between the new state at filling factor $v > +1$ and the more familiar state at lower densities. By tracking the dependence of this boundary on temperature and magnetic field, we find that it exhibits an electronic analogue[29–32] of the famous Pomeranchuk effect[1] in $^3$He. In that system, a transition from a Fermi liquid to a solid occurs upon increasing temperature, driven by the high spin entropy of the localized atoms in the solid. Similarly, we find that the new state above $v = +1$ is favored relative to the metallic state at $v < +1$ upon raising the temperature, due the former's high magnetic entropy. The transition near $v = +1$ can also be driven by an in-plane magnetic field, due to the energy gain associated with polarizing the free moments. (A related effect near $v = -1$ was proposed very recently, on the basis of transport measurements[33].) The existence of the hybrid state we observe here, with its itinerant electrons coexisting with strongly fluctuating magnetic moments, calls for a new understanding of electron correlations in MATBG.

The data reported here is measured using two independent techniques on two conceptually different devices. The bulk of the results are obtained from local measurements of the electronic entropy[34,35] and compressibility using a scanning nanotube single-electron transistor (SET) on hBN-encapsulated twisted bilayer device (Device 1, Fig. 1a). We focus on a spatial region whose twist angle is close to the theoretical magic angle, and is homogenous over a large area (5μm × 4μm) to within the third digit, $\theta = 1.130 \pm 0.005$. Similar results are obtained from global entropy measurements using a monolayer graphene sensor to detect the chemical potential of MATBG (Device 2, Fig. 3a). Both methods have been described elsewhere[21,36].

The inverse compressibility, $d\mu/dn$, measured in Device 1 at $T = 15$K as a function of the filling factor, $\nu = n/(n_s/4)$ (where $n_s$ corresponds to four electrons per moiré unit cell), is shown in Fig. 1b. As reported previously[21], sharp jumps in $d\mu/dn$ are observed close to integer $\nu$'s, where the system rapidly evolves from high to low compressibility, reflecting an abrupt reconstruction of the Fermi surface. These were termed Dirac revivals since they were interpreted as resets of partially filled energy bands back to near charge neutrality, leading to the decreased compressibility. As seen in the figure, the cascade of revivals is already very prominent at these relatively high temperatures (i.e. above typical correlated insulator and superconducting critical temperature scales). Measurements of $\rho_{xx}$ vs. $\nu$ at various temperatures (Fig. 1c) show insulating behavior at $\nu = 2$ and semi-metallic behavior at $\nu = 0$. As previously-noted[37], $\rho_{xx}$ shows a step-like increase across $\nu \approx 1$ at high temperatures, and this feature gradually disappears with lowering the temperature (although a small peak in the resistivity close to $\nu = 1$ remains). This behavior is different from the insulating behavior observed at other integer filling factors.

The unusual physics of the electronic state around $\nu = 1$ is revealed by measuring the dependence of the inverse compressibility, $d\mu/dn$, on temperature, $T$, and parallel magnetic field, $B_\parallel$. In Fig. 2a we examine the temperature dependence of $d\mu/dn$ near $\nu = 1$ at $B_\parallel = 0$T. At low temperature, a jump[21] in $d\mu/dn$ occurs at a filling factor slightly larger than 1. Increasing the temperature moves the jump towards a lower filling factor, and surprisingly, increases the magnitude of the jump, rather than smearing it. A similar measurement with $B_\parallel = 12$T is shown in Fig. 2b. Compared to $B_\parallel = 0$T, at low $T$ the jump is much larger and closer to $\nu = 1$. Increasing the temperature at $B_\parallel = 12$T maintains the jump close to $\nu = 1$ and, oppositely to the $B_\parallel = 0$T case, reduces its amplitude and increases its width.

The chemical potential, $\mu(\nu, T)$ (measured relative to that at the charge neutrality point), can be obtained by integrating $d\mu/dn$ over the density at different temperatures (Fig. 2c,d). We see that $\mu$ has a strong temperature dependence for a certain range of $\nu$'s. This is clearly seen when we plot $\mu$ vs. $T$ at two representative $\nu$'s (Fig. 2c, inset). At $\nu = 0.2$, $\mu$ is practically independent of $T$ (blue). In contrast, at $\nu = 0.9$ (red) we see that $\mu$ is nearly constant until $T \sim 4K$, after which it starts to decrease approximately linearly with $T$. At $\nu > 1.15$, $\mu$ is again nearly temperature independent. Comparing $\mu$ at $B_\parallel = 0T$ (Fig. 2c) and $B_\parallel = 12T$ (Fig. 2d) reveals a clear contrast: whereas for $B_\parallel = 0T$, $\mu$ is a decreasing function of temperature for $0.4 < \nu < 1.15$, for $B_\parallel = 12T$, $\mu$ decreases with $T$ for $\nu < 0.9$ and increases for $\nu > 0.9$.

Measurements of the temperature dependence of $\mu$ allow us to directly determine the entropy of the system, by integrating Maxwell's relation: $\left(\frac{\partial s}{\partial \nu}\right)_T = -\left(\frac{\partial \mu}{\partial T}\right)_\nu$, to obtain $s(\nu, T)$ (where $s$ is the entropy per moiré unit cell). For more details on the procedure of extracting the entropy, see Supplementary Information section SI1. Fig. 2e shows $s(\nu)$ at $T \approx 10K$ (obtained by extracting the slope of $\mu$ vs. $T$ in the range $T = 4.5K - 15K$), for $B_\parallel = 0T, 4T, 8T,$ and $12T$. At $B_\parallel = 0T$ the entropy is small at low $\nu$'s, climbs close to $\nu = 1$, and remains roughly constant between $\nu = 1$ and $2$ at about $1.2 k_B$ per unit cell. Near $\nu = 2$ the entropy has a sharp drop, and it starts decreasing towards zero after $\nu = 3$. We note that the dependence of the entropy on $\nu$ is qualitatively different from that of the compressibility. Specifically, whereas the compressibility drops sharply near $\nu = 1$ (Fig. 2a), the entropy does not drop, but rather remains at a high value.

An important insight into the origin of this large entropy can be gained by examining its magnetic field dependence. As seen in Fig. 2e, the entropy above $\nu = 1$ depends strongly on $B_\parallel$. In particular, at $B_\parallel = 12T$, most of the entropy between $\nu = 1$ and $2$ is quenched. The inset shows $s(B_\parallel = 0T) - s(B_\parallel = 12T)$ vs. $\nu$ (with errorbars indicated by the purple shading; see Supplementary Information SI1). The entropy difference increases sharply near $\nu = 1$, reaching a maximum of $0.9 \pm 0.1 k_B$ between $\nu = 1$ and $2$. To appreciate the significance of this value, recall that an entropy of $k_B \ln(2) \approx 0.7 k_B$ corresponds to two degenerate states on each moiré unit cell. Moreover, in a Fermi liquid, we would expect a much weaker change of the entropy with $B_\parallel$ (Supplementary

Information SI4), of the order of $k_B$ times the ratio of the Zeeman energy (about 1meV at $B_\parallel = 12$T) to the bandwidth, estimated to be $W \sim 30$meV (see below). Finally, we observe that the entropy vs. $\nu$ at $B_\parallel = 12$T shows a cascade of drops following each integer $\nu$, similar to the cascade of revivals seen in the compressibility (Supplementary Info. SI5), as obtained in a mean-field calculation (Supplementary Info. SI3). The dramatic quenching of entropy by a moderate in-plane magnetic field is strongly suggestive of its magnetic origin.

To test the robustness of our results, we measured the entropy in a completely different setup, in which a sheet of monolayer graphene is used to sense the chemical potential of the MATBG, averaged over the entire device[36] (Fig. 3a). Fig. 3b shows the entropy extracted in three different temperature ranges. In particular, we see that the entropy from global measurements done in the range $T = 4$K $- 16$K (blue curve, Fig. 3b) is in good agreement with the locally measured one over a similar range of temperatures (see inset for a comparison). The agreement is both in the overall shape and magnitude of $s(\nu)$, and in detailed features, such as the step-like drop near $\nu = 2$. The global measurement setup further extends the experiment to the high temperature regime. As seen in Fig. 3b, upon raising the temperature, the minimum in the entropy at $\nu = 0$ gradually fills in, evolving from a double-dome structure at low $T$ (corresponding to the valence and conduction flat bands) to a single dome at high $T$. This high temperature dependence can be qualitatively reproduced by a naïve calculation for a system of non-interacting electrons in Dirac bands, whose density of states rises linearly from the charge neutrality point until the band edges (Fig. 3c). The merging of the domes in $s(\nu)$ occurs when the temperature exceeds some fraction of the bandwidth. Calibrating the bandwidth in the calculation using the measured entropy at $T \approx 55$K gives $W \approx 30$meV (where $W$ is the full bandwidth – from valence band bottom to conduction band top), in rough agreement with the values deduced from STM[16–19] and compressibility[36] experiments. Of course, we do not expect the free-electron picture to apply at low temperatures, since there interactions change the physics dramatically. We note that the measured $s(v)$ in the valence band (Fig. 3b) is approximately a mirror image of $s(v)$ in the conduction band, although it is somewhat smaller and has less pronounced features. This is consistent with the fact that the observed cascade of revivals in the inverse compressibility is weaker in the valence band relative to that in the conduction band[21,36].

Our results so far show that a dramatic change occurs in the compressibility and the entropy near $v = 1$. The compressibility experiences an abrupt drop at the revival transition (Fig. 2a,b), and at approximately the same filling, the magnetic-field-dependent part of the entropy sharply rises (Fig. 2e, inset). This rapid change may be due to a continuous buildup of electronic correlations. Alternatively, it can be interpreted as an underlying first-order phase transition between two distinct phases. Naively, one would then expect a discontinuous jump in thermodynamic properties and hysteretic behavior across the transition, which are not observed. However, we note that a true first-order phase transition can never occur in two dimensions in the presence of disorder or long-range Coulomb interactions[38], as these will always broaden the transition into a mesoscale coexistence region. Experimentally, since the observed compressibility revival feature is relatively sharp, it can be precisely tracked, mapping a phase diagram as the function of temperature and magnetic field. Below, we show that interpreting this revival feature as a proxy for a first-order transition naturally explains much of the underlying physics.

We define the filling factor $v_R$ of the revival feature as the midpoint of the sharp rise in $d\mu/dn$ (tracking the beginning or the end of the rise leads to similar conclusions, see Supplementary Info. SI5). As we have seen in Fig. 2a, raising the temperature leads to an observable change in $v_R$. A similar measurement of $d\mu/dn$ vs. $v$ at different magnetic fields (Fig. 4a) shows that $v_R$ also shifts with $B_\parallel$. The measured locations of the revival feature as a function of $B_\parallel$ and $T$ form a surface in the $(v, B_\parallel, T)$ space, shown in Fig. 4b. Projections of this surface onto the $(v, B_\parallel)$ and $(v, T)$ planes are presented in Figs. 4c,d. Examining the magnetic field dependence of $v_R$ (Fig. 4c), we see that at $T = 2.8$K, $v_R$ is weakly dependent on $B_\parallel$ at low fields, but starts to decrease linearly with field above $B_\parallel \approx$ 4T. At higher temperatures, $v_R$ is similarly insensitive to magnetic field at low $B_\parallel$, and decreases with increasing $B_\parallel$ at higher fields. The crossover field between the two behaviors increases as the temperature increases.

Another interesting aspect of the evolution of $v_R$ is highlighted by looking at its temperature dependence at the different magnetic fields (Fig. 4d). At $B_\parallel = 0\text{T}$ (blue) $v_R$ is linear in $T$ at low temperatures, and curves up at higher temperatures. As the magnetic field increases, the entire curve shifts towards smaller $v$'s, and simultaneously its slope at low temperatures changes sign. At the highest field, $B_\parallel = 12\text{T}$, $v_R$ first increases with temperature, reaches a maximum at $T \approx 9\text{K}$, and then decreases at a higher $T$.

The phenomenology seen in Figs. 4b-d can be understood within a simple interpretation, in terms of a first-order phase transition at $v = v_R$ between two phases: a Fermi liquid phase below $v_R$, and a 'free moment' phase above it. The latter phase has a high concentration of free moments (of the order of one per moiré site), coexisting with a low density of itinerant electrons. Within this framework, the movement of the transition point $v_R$ as a function of $B_\parallel$ and $T$ reflects the magnetization and entropy differences between the two neighboring phases.

At $B_\parallel = 0\text{T}$, the free moment phase has a higher entropy than the Fermi liquid, due to thermal fluctuations of the moments. Hence, the former phase becomes increasingly entropically-favorable the higher the temperature. This explains the observed decrease of $v_R$ with increasing $T$ at low fields (Fig. 4d). Raising the temperature at a fixed $v$ may therefore drive a transition from the Fermi liquid to the free moments phase, an electronic analogue of the Pomeranchuk effect. As $B_\parallel$ increases and the Zeeman energy exceeds the temperature, the moments become nearly fully polarized and their entropy is quenched (as is observed directly in Fig. 2e). Consequently, one expects that at low temperatures and sufficiently high fields, the Fermi liquid phase would be favored by raising the temperature, but that this trend will reverse once the temperature becomes larger than the Zeeman energy. This explains the non-monotonic behavior of $v_R$ as a function of $T$, seen at $B_\parallel = 12\text{T}$ in Fig. 4d. The main features of the phase boundary are qualitatively reproduced in an explicit model for the thermodynamics of the two phases (see Supplementary Info. SI7), as shown in the insets of Figs. 4b,c,d. We note that, although this simple model of a transition between a Fermi liquid and a localized moment phase explains much of the phenomenology present in Fig. 4, a complete understanding requires a more detailed consideration of interactions also for $\nu < \nu_R$ (e.g. to account for negative compressibility, revival strength, etc)[21,36].

To conclude, our measurements reveal the emergence of a large entropy above $\nu = 1$. This excess entropy is suppressed upon the application of a magnetic field, highlighting its magnetic origin. We observe two regions with vastly different properties on either side of $\nu = 1$. We map the boundary between these two regions, and show that it can be naturally understood as a transition from a Fermi liquid-like phase to a free moment phase. The latter has several highly unusual characteristics: its moments remain strongly fluctuating down to very low temperatures (3K or less), much lower than the electronic bandwidth and the onset temperature of the revival feature in the compressibility. Moreover, this phase is compressible and metallic, with no sign of a thermodynamic gap. These observations raise important challenges for our understanding of electron correlations in MATBG. In particular, what is the origin of the soft magnetic excitations? Such soft collective modes have been predicted in insulating states of MATBG[25–27]. However, our experiments do not exhibit insulating behavior near $\nu = 1$. One can imagine such soft modes emerging also in a gapless state, either as low-energy magnetic collective excitations of itinerant electrons, or due to a partial localization of some of the electrons, coexisting with other itinerant ones. The existence of such free moments at low temperatures should have far-reaching consequences for the transport behavior and the phase diagram of this fascinating strongly correlated electron system.

**References**

1. Pomeranchuk, I. On the thery of He3. *Zh.Eksp.Teor.Fiz* **20**, 919 (1950).

2. Bistritzer, R. & MacDonald, A. H. Moiré bands in twisted double-layer graphene. *Proc. Natl. Acad. Sci.* **108**, 12233–12237 (2011).

3. Cao, Y. *et al.* Correlated insulator behaviour at half-filling in magic-angle graphene superlattices. *Nature* **556**, 80–84 (2018).

4. Cao, Y. *et al.* Unconventional superconductivity in magic-angle graphene superlattices. *Nature* **556**, 43–50 (2018).

5. Li, G. *et al.* Observation of Van Hove singularities in twisted graphene layers. *Nat. Phys.* **6**, 109–113 (2010).

6. Suárez Morell, E., Correa, J. D., Vargas, P., Pacheco, M. & Barticevic, Z. Flat bands in slightly twisted bilayer graphene: Tight-binding calculations. *Phys. Rev. B* **82**, 121407 (2010).

7. Regan, E. C. *et al.* Mott and generalized Wigner crystal states in $WSe_2/WS_2$ moiré superlattices. *Nature* **579**, 359–363 (2020).

8. Tang, Y. *et al.* Simulation of Hubbard model physics in $WSe_2/WS_2$ moiré superlattices. *Nature* **579**, 353–358 (2020).

9. Yankowitz, M. *et al.* Tuning superconductivity in twisted bilayer graphene. *Science* **363**, 1059–1064 (2019).

10. Lu, X. *et al.* Superconductors, orbital magnets and correlated states in magic-angle bilayer graphene. *Nature* **574**, 653–657 (2019).

11. Nuckolls, K. P. *et al.* Strongly Correlated Chern Insulators in Magic-Angle Twisted Bilayer Graphene. *arXiv* 2007.03810 (2020).

12. Wu, S., Zhang, Z., Watanabe, K., Taniguchi, T. & Andrei, E. Y. Chern Insulators and Topological Flat-bands in Magic-angle Twisted Bilayer Graphene. *ArXiv* 2007.03725 (2020).

13. Das, I. *et al.* Symmetry broken Chern insulators and magic series of Rashba-like

Landau level crossings in magic angle bilayer graphene. *Arxiv* 2007.13390 (2020).

14. Sharpe, A. L. *et al.* Emergent ferromagnetism near three-quarters filling in twisted bilayer graphene. *Science* **365**, 605–608 (2019).

15. Serlin, M. *et al.* Intrinsic quantized anomalous Hall effect in a moiré heterostructure. *Science* **367**, 900–903 (2020).

16. Kerelsky, A. *et al.* Maximized electron interactions at the magic angle in twisted bilayer graphene. *Nature* **572**, 95–100 (2019).

17. Xie, Y. *et al.* Spectroscopic signatures of many-body correlations in magic-angle twisted bilayer graphene. *Nature* **572**, 101–105 (2019).

18. Jiang, Y. *et al.* Charge order and broken rotational symmetry in magic-angle twisted bilayer graphene. *Nature* **573**, 91–95 (2019).

19. Choi, Y. *et al.* Electronic correlations in twisted bilayer graphene near the magic angle. *Nat. Phys.* **15**, 1174–1180 (2019).

20. Tomarken, S. L. *et al.* Electronic Compressibility of Magic-Angle Graphene Superlattices. *Phys. Rev. Lett.* **123**, 046601 (2019).

21. Zondiner, U. *et al.* Cascade of phase transitions and Dirac revivals in magic-angle graphene. *Nature* **582**, 203–208 (2020).

22. Po, H. C., Zou, L., Vishwanath, A. & Senthil, T. Origin of Mott insulating behavior and superconductivity in twisted bilayer graphene. *Phys. Rev. X* **8**, 031089 (2018).

23. Song, Z. *et al.* All Magic Angles in Twisted Bilayer Graphene are Topological. *Phys. Rev. Lett.* **123**, 036401 (2019).

24. Ahn, J., Park, S. & Yang, B.-J. Failure of Nielsen-Ninomiya Theorem and Fragile Topology in Two-Dimensional Systems with Space-Time Inversion Symmetry: Application to Twisted Bilayer Graphene at Magic Angle. *Phys. Rev. X* **9**, 021013 (2019).

25. Bultinck, N. *et al.* Ground State and Hidden Symmetry of Magic-Angle Graphene at Even Integer Filling. *Phys. Rev. X* **10**, 031034 (2020).

26. MacDonald, A. H. private communication.

27. Wu, F. & Das Sarma, S. Collective Excitations of Quantum Anomalous Hall Ferromagnets in Twisted Bilayer Graphene. *Phys. Rev. Lett.* **124**, 046403 (2020).

28. Wong, D. *et al.* Cascade of electronic transitions in magic-angle twisted bilayer graphene. *Nature* **582**, 198–202 (2020).

29. McWhan, D. B. *et al.* Electronic Specific Heat of Metallic Ti-Doped $V_2O_3$. *Phys. Rev. Lett.* **27**, 941–943 (1971).

30. Spivak, B. & Kivelson, S. A. Phases intermediate between a two-dimensional electron liquid and Wigner crystal. *Phys. Rev. B* **70**, 155114 (2004).

31. Continentino, M. A., Ferreira, A. S., Pagliuso, P. G., Rettori, C. & Sarrao, J. L. Solid state Pomeranchuk effect. *Phys. B Condens. Matter* **359–361**, 744–746 (2005).

32. Pustogow, A. *et al.* Quantum spin liquids unveil the genuine Mott state. *Nat. Mater.* **17**, 773–777 (2018).

33. Saito, Y. *et al.* Isospin Pomeranchuk effect and the entropy of collective excitations in twisted bilayer graphene. *ArXiv* 2008.10830 (2020).

34. Kuntsevich, A. Y., Tupikov, Y. V, Pudalov, V. M. & Burmistrov, I. S. Strongly correlated two-dimensional plasma explored from entropy measurements. *Nat. Commun.* **6**, 7298 (2015).

35. Hartman, N. *et al.* Direct entropy measurement in a mesoscopic quantum system. *Nat. Phys.* **14**, 1083–1086 (2018).

36. Park, J. M., Cao, Y., Watanabe, K., Taniguchi, T. & Jarillo-Herrero, P. Flavour Hund's Coupling, Correlated Chern Gaps, and Diffusivity in Moiré Flat Bands. *Arxiv* 2008.12296 (2020).

37. Chen, S. *et al.* Electrically tunable correlated and topological states in twisted monolayer-bilayer graphene. *Arxiv* 2004.11340 (2020).

38. Spivak, B. & Kivelson, S. A. Transport in two dimensional electronic micro-emulsions. *Ann. Phys.* **321**, 2071–2115 (2006).

39. Uri, A. *et al.* Mapping the twist-angle disorder and Landau levels in magic-angle graphene. *Nature* **581**, 47–52 (2020).

**Acknowledgements:** We thank Ehud Altman, Eva Andrei, Eslam Khalaf, Steve Kivelson, Sankar Das Sarma, Gal Shavit, Joey Sulpizio, Senthil Todadri, Aviram Uri, Ashvin Vishwanath, Michael Zaletel and Eli Zeldov for useful suggestions. E.B. is grateful to Andrea Young for drawing his attention to the unusual physics near $\nu = \pm 1$, sharing his unpublished data, and for a collaboration on a related experimental and theoretical work[33], proposing that a similar effect to the one discussed here occurs near $\nu = -1$, based on transport measurements. In this work, in contrast, we measured the entropy directly, and mapped the entire phase diagram near $\nu = 1$ using compressibility measurements. Work at Weizmann was supported by the Leona M. and Harry B. Helmsley Charitable Trust grant, ISF grants (712539 & 13335/16), Deloro award, Sagol Weizmann-MIT Bridge program, the ERC-Cog (See-1D-Qmatter, no. 647413), the ISF Research Grants in Quantum Technologies and Science Program (994/19 & 2074/19), the DFG (CRC/Transregio 183), ERC-Cog (HQMAT, no. 817799), EU Horizon 2020 (LEGOTOP 788715) and the Binational Science Foundation (NSF/BMR-BSF grant 2018643). Work at MIT was primarily supported by the US Department of Energy (DOE), Office of Basic Energy Sciences (BES), Division of Materials Sciences and Engineering under Award DE-SC0001819 (J.M.P.). Help with transport measurements and data analysis were supported by the National Science Foundation (DMR-1809802), and the STC Center for Integrated Quantum Materials (NSF Grant No. DMR-1231319) (Y.C.). P.J-H acknowledges support from the Gordon and Betty Moore Foundation's EPiQS Initiative through Grant GBMF9643. The development of new nanofabrication and characterization techniques enabling this work has been supported by the US DOE Office of Science, BES, under award DE-SC0019300. K.W. and T.T. acknowledge support from the Elemental Strategy Initiative conducted by the MEXT, Japan, Grant Number JPMXP0112101001, JSPS KAKENHI Grant Numbers JP20H00354 and the CREST(JPMJCR15F3), JST. This work made use of the Materials Research Science and Engineering Center Shared Experimental Facilities supported by the National Science Foundation (DMR-0819762) and of Harvard's Center for Nanoscale Systems, supported by the NSF (ECS-0335765).

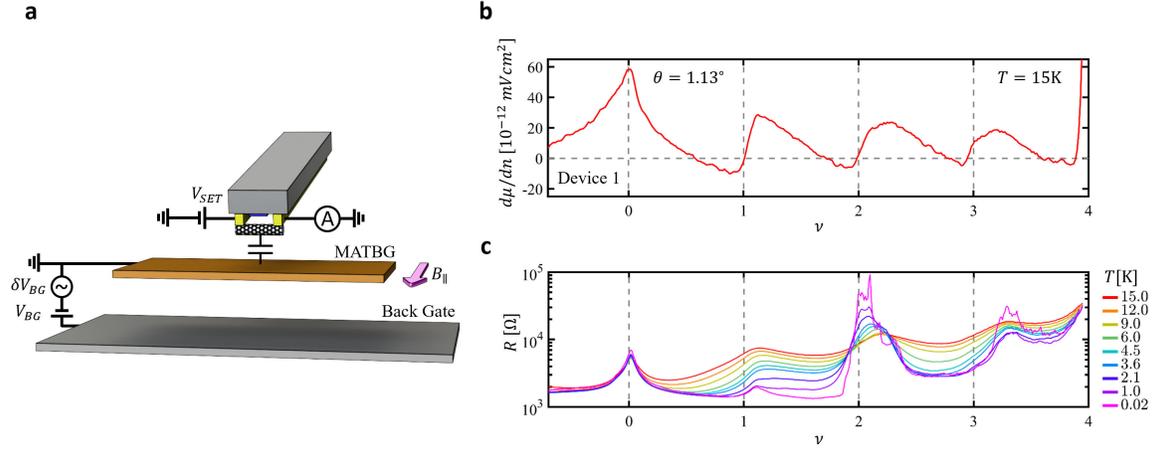

**Figure 1: Experimental setup and device characterization. a.** A nanotube-based single electron transistor (SET) is used to measure the local electronic compressibility and entropy of magic angle twisted bilayer graphene (MATBG). The MATBG is encapsulated between top and bottom h-BN layers (not shown) and has a metallic back-gate. By monitoring the current through the SET, we track changes in the MATBG chemical potential, $d\mu$, in response to a density modulation, $dn$, produced by an a.c. voltage on the back-gate[21], $\delta V_{BG}$. A d.c. back-gate voltage, $V_{BG}$, sets the overall carrier density in the MATBG, $n$. Some of the measurements are performed in a parallel magnetic field, $B_\parallel$ (indicated). **b.** Inverse compressibility, $d\mu/dn$, measured as a function of the moiré lattice filling factor, $\nu = n/(n_s/4)$, at $T = 15K$ ($n_s$ is the density that correspond to 4 electrons per moiré site). Measurements are done on a large spatial domain ($\sim 5\mu m \times 4\mu m$) throughout which the twist angle is extremely homogenous, $\theta = 1.130° \pm 0.005$ (measured by spatial mapping of the $V_{BG}$ that corresponds to $n_s$, as in Refs. [21,39]). As seen previously[21], a jump of $d\mu/dn$ appears near all integer filling factors. This jump corresponds to a Fermi surface reconstruction, in which some combination of the spin/valley flavors filling is reset back to near the charge neutrality point, and correspondingly $d\mu/dn$ shows a cascade of sawtooth features as a function of density. The trace is measured at $T = 15K$, showing that even at this high temperature this sawtooth cascade is well developed **c.** Two-probe resistance, $R$, measured as a function of $\nu$ and temperature. Notice that unlike the inverse compressibility, which measures a local quantity, the resistance gives an averaged result over domains with different twist angle. Therefore, the resistance maxima are slightly shifted from the usual integer $\nu$ values, probably because another domain with a small difference in twist angle dominates the transport characteristics globally.

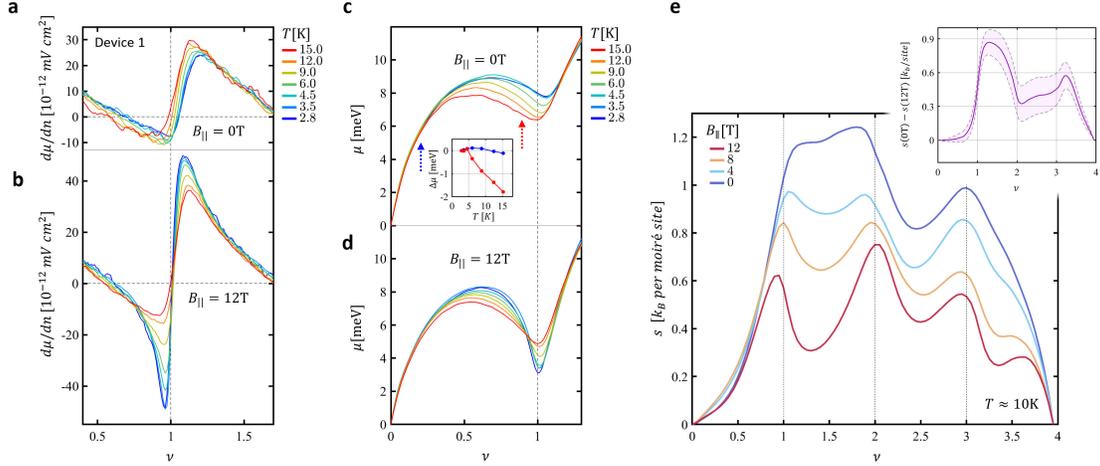

**Figure 2: Measurement of large magnetic entropy above $\nu = 1$**. **a.** Inverse compressibility, $d\mu/dn$, as a function of $\nu$, near $\nu = 1$, measured at zero parallel magnetic field, $B_\parallel = 0T$, and at several temperatures. With increasing $T$, the jump in $d\mu/dn$ moves toward lower $\nu$ and becomes stronger. **b.** Same measurement done at $B_\parallel \approx 12T$. Here, opposite to the zero-field case, increasing $T$ reduces the magnitude of the $d\mu/dn$ jump, as expected from thermal smearing. **c.** The chemical potential $\mu(\nu)$ (relative to that of the charge neutrality point) at $B_\parallel = 0T$, obtained by integrating the $d\mu/dn$ signal in panel a with respect to $n$. Inset: $\mu(T, \nu) - \mu(T = 2.8K, \nu)$ for $\nu = 0.2$ (blue) and $\nu = 0.9$ (red). At $\nu = 0.2$ the chemical potential is nearly temperature independent, whereas at $\nu = 0.9$ it is roughly constant until $T \sim 4K$ and then start decreasing approximately linearly with $T$. **d.** Similar to c, but at $B_\parallel = 12T$. In contrast to the zero-field case, here, below $\nu \approx 0.9$, $\mu$ decreases with $T$ while above $\nu \approx 0.9$ $\mu$ increases with $T$. **e.** The electronic entropy in units of $k_B$ per moiré unit cell, as a function of $\nu$ at $T \approx 10K$ and at various parallel magnetic fields, $B_\parallel = 0,4,8,12T$. To obtain the entropy we determine the partial derivative $(\partial \mu/\partial T)_{\nu,B_\parallel}$ from a linear fit to the measured $\mu$ vs. $T$ in the range $T = 4.5K - 15K$. The entropy per moiré cell is then obtained by integrating Maxwell's relation: $(\partial s/\partial \nu)_{T,B_\parallel} = -(\partial \mu/\partial T)_{\nu,B_\parallel}$, over $\nu$ (see Supp Info. for details) . At $B_\parallel = 0$ the entropy climbs rapidly near $\nu = 1$ to a value of $1.2 k_B$ per moiré cell. Inset: the difference between the entropies at low and high fields, $s(B_\parallel = 0T) - s(B_\parallel = 12T)$. The purple shading shows the estimated error bar.

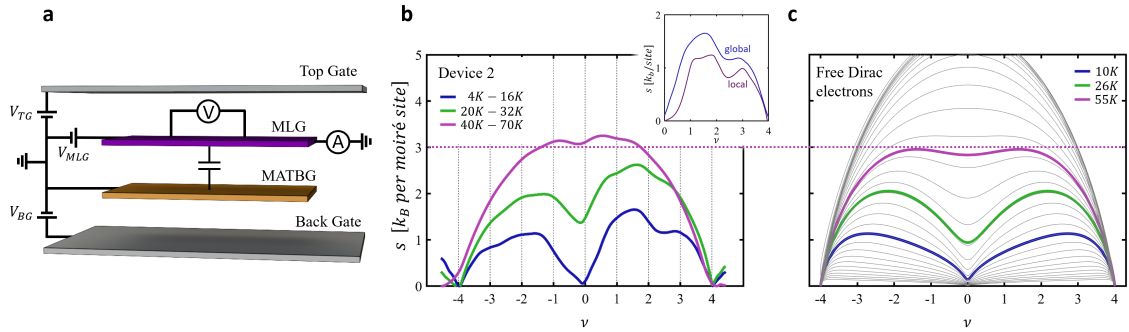

**Figure 3: Temperature dependence of the Entropy. a.** Experimental setup for measuring the global entropy, averaged over the entire device[36]. The device consists of MATBG and a monolayer graphene (MLG) sensor layer, separated by an ultrathin (1 $nm$) layer of h-BN (not shown), as well as top and bottom metallic gates. By balancing the electrochemical potential of the adjacent layers in the device, we can obtain the relationship between the density and chemical potential of MATBG and MLG and the gate voltages applied to the system. In the special case where the density of MLG is zero, i.e. at its charge neutrality point, the chemical potential of MATBG is directly proportional to the voltage applied to the top gate. This technique allows us to reliably extract the chemical potential and entropy of MATBG at temperatures up to 70 K. **b.** The measured entropy, in units of $k_B$ per moiré unit cell, as a function of $\nu$ at three different temperature ranges (top legend). The entropy derivative, $ds/d\nu$, is obtained from a linear fit to $\mu$ vs. $T$ in the corresponding temperature range, and is then integrated over $\nu$ to yield the entropy per moiré unit cell (similar to Fig. 2e). Inset: comparison between the $\nu$ dependences of the entropies, measured at the low temperature range, obtained from local and global measurements. **c.** The entropy as a function of $\nu$ and $T$ calculated for a system of four degenerate non-interacting Dirac bands (whose density of states climbs linearly with energy from the Dirac point to the end of the conduction or the valence band). The color-coded lines show the curves whose temperatures correspond to the mean of the temperature ranges of the experimental curves. The gray lines represent the entire evolution from zero temperature to high temperature, where the entropy saturates on a value of $8ln(2) \approx 5.5$, where the factor 8 reflects the total number of energy bands. A bandwidth of $W = 30meV$ is chosen such that the calculated value of the entropy at the highest temperature roughly matches the one obtained from the measured curve at the same temperature.

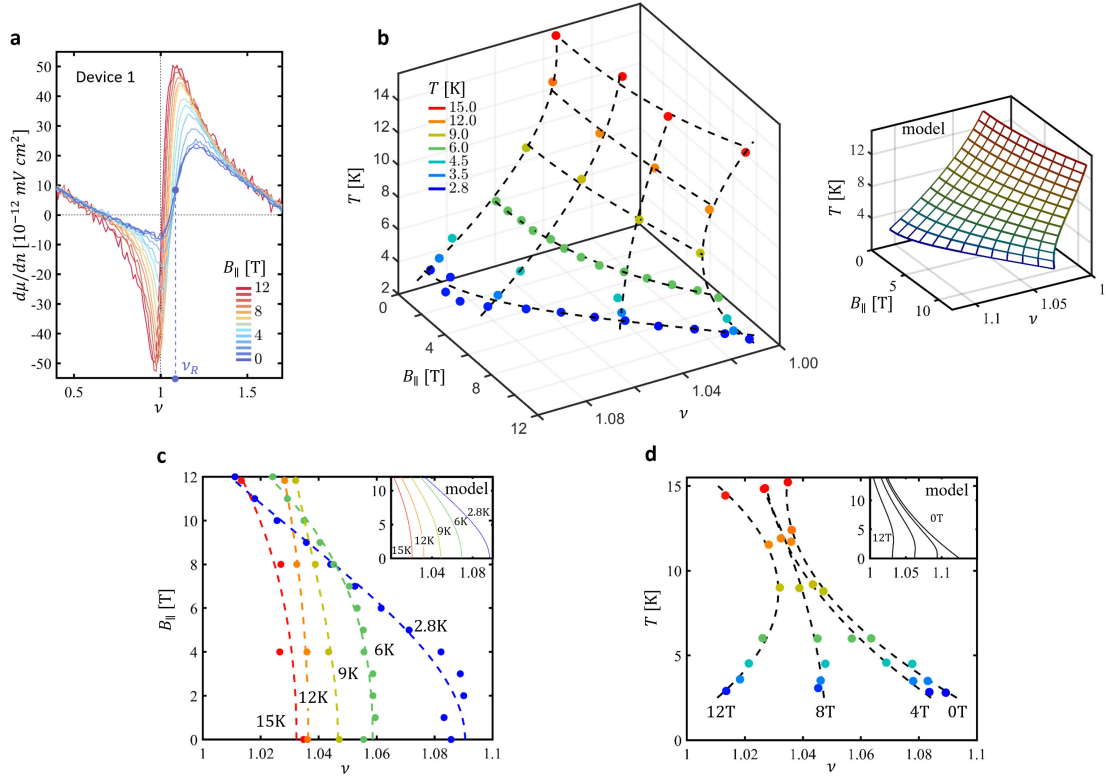

**Figure 4: Experimental phase diagram. a.** The inverse compressibility, $d\mu/dn$, measured as a function of $\nu$ near $\nu = 1$, at several values of parallel magnetic field, $B_\parallel$. We track the filling factor that corresponds to the center the jump in $d\mu/dn$ (labeled $\nu_R$). Visibly, the application of $B_\parallel$ pushes $\nu_R$ to lower values. **b.** Measured $\nu_R$ as a function of $B_\parallel$ and $T$, plotted as dots in the $(\nu, B_\parallel, T)$ space (the dots are colored by their temperature). The dashed lines are polynomial fits to the dots at constant $B_\parallel$ or constant $T$. Inset: the same surface calculated from a simple model that assumes a transition between a Fermi liquid and a metallic phase that contains one free moments per moiré site (see text). **c.** Projection of the data in panel b onto the $(\nu, B_\parallel)$ plane, showing the dependence of $\nu_R$ on $B_\parallel$ for various temperatures. At low fields, $\nu_R$ is independent of field but it becomes linear in $B_\parallel$ at high fields, a behavior expected from the field polarization of free moments (see text). Inset: curves calculated from the model. **d.** Projection onto the $(\nu, T)$ plane, showing the dependence of $\nu_R$ on $T$ for various magnetic fields. At $B_\parallel = 0T$, $\nu_R$ is linear in $T$ at small $T$'s and then curves up at higher $T$'s. At high magnetic field, the dependence of $\nu_R$ on $T$ becomes non-monotonic. Inset: curves calculated from the model.

Supplementary materials for:

# Entropic evidence for a Pomeranchuk effect in Magic Angle graphene

Asaf Rozen[†], Jeong Min Park[†], Uri Zondiner[†], Yuan Cao[†], Daniel Rodan-Legrain, Takashi Taniguchi, Kenji Watanabe, Yuval Oreg, Ady Stern, Erez Berg[*], P. Jarillo-Herrero [*] and Shahal Ilani[*]

## Contents

## SI1. Extraction of the entropy

In both the local and global measurements, we determine the entropy using a Maxwell relation, relating the partial derivatives of the entropy with respect to the filling factor to that of the chemical potential with respect to temperature:

$$(\partial s/\partial \nu)_{T,B_\parallel} = -(\partial \mu/\partial T)_{B_\parallel,\nu}$$

where $s$ is the entropy per moiré unit cell. In the global measurements, we probe the chemical potential of the MATBG directly using a monolayer graphene sensor. The measurement determines the chemical potential relative to that at the charge neutrality point (CNP):

$$\Delta\mu(\nu, T, B_\parallel) = \mu(\nu, T, B_\parallel) - \mu_{CNP}(T, B_\parallel).$$

In the local measurements, we use a nanotube single electron transistor to measure the inverse compressibility and integrate it over the density, to obtain the same quantity:

$$\Delta\mu(\nu, T, B_\parallel) = \mu(\nu, T, B_\parallel) - \mu_{CNP}(T, B_\parallel) = \int_0^n (\partial\mu/\partial n)_{B_\parallel,T}\, dn'.$$

In these measurements, the inverse compressibility is probed at typical frequencies of few hundred Hz, and with an excitation $\delta V_{BG} = 40 mV$ on the back gate, chosen to be small enough as to not smear essential features.

The entropy then follows from:

$$s(\nu, T, B_\parallel) = \int_0^\nu (\partial s/\partial \nu)_{T,B_\parallel}\, d\nu' = -\int_0^\nu (\partial\mu/\partial T)_{B_\parallel,\nu}\, d\nu'$$

$$= -\int_0^\nu \frac{d(\Delta\mu)}{dT} d\nu' - \int_0^\nu \frac{d\mu_{CNP}}{dT} d\nu'$$

The first term provides the $\nu$-dependent part of the entropy. The second one, which we do not measure directly, adds a linear term in $\nu$. The value of this constant is determined by making the assumption that inside the gap separating the conduction flat band and the higher dispersive band, namely at $\nu = 4$, the electronic entropy is zero. To see why this assumption is justified we note that inside a gap, the electronic entropy is given by $s = 16 k_B \frac{E_g}{W} e^{-\frac{E_g}{2k_B T}}$ (where $W$ is the width of the flat band, and $E_g$ is the size of the gap to the dispersive band). Our compressibility measures directly the size of the gap to be $E_g \approx 30 meV$, and estimate the bandwidth to be of similar magnitude $W \approx 30 - 40 meV$. The entropy in such gap at $T \approx 10K$ is $s \approx 4 \cdot 10^{-7} k_B$, making our assumption well justified for the relevant temperatures reported in the paper.

Fig. S1a shows the derivative of the entropy per electron with respect to $\nu$ for three different temperature ranges, from the measurements done in Device 2. using the global measurements. We removed a constant background in $ds/d\nu$ ($\nu$) to account for the variation of $\mu$ with $T$ at charge neutrality, such that the entropy at $\nu = \pm 4$ is zero. For each temperature range, $\mu$ was assumed to be linearly dependent on T at a given $\nu$. The confidence bound of 95% is shown for this linear fitting process. The entropy obtained after integration is shown in Fig. S1b. The error highlighted bands show the propagated uncertainty in this integration process.

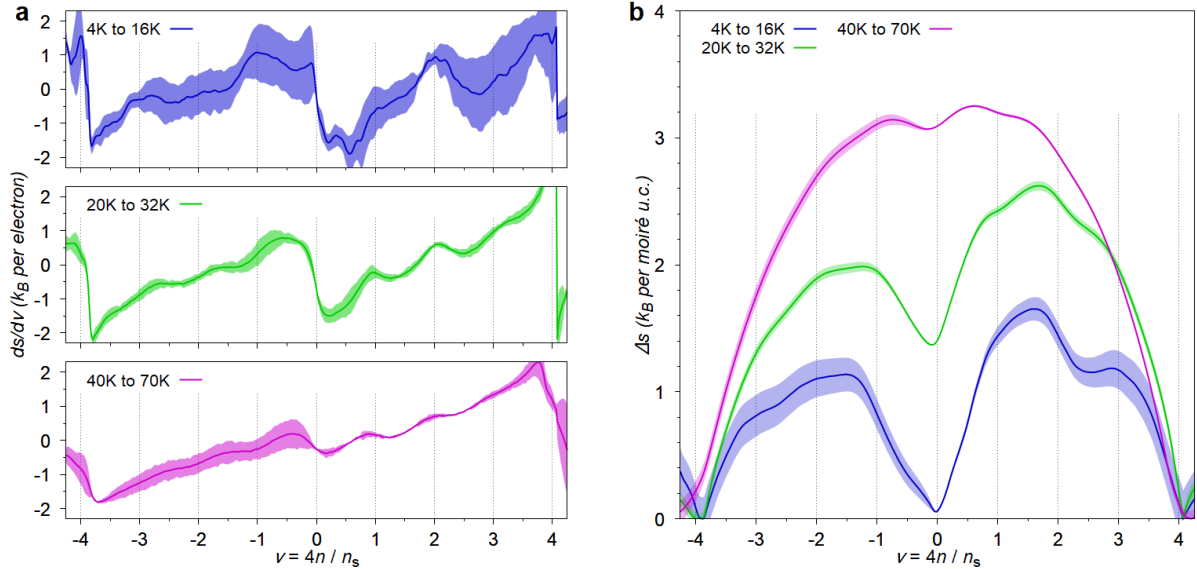

**Fig. S1: Extraction of entropy in Device 2 for different temperatures**. (a) Derivative of entropy with respect to $\nu$ obtained from Maxwell's relation $(\partial s/\partial \nu)_{T,B_\parallel} = -(\partial \mu/\partial T)_{B_\parallel,\nu}$ for three temperature ranges spanning 4 K to 70 K. (b) Entropy change $\Delta s$ per moiré unit cell with respect to the band insulators at $\nu = \pm 4$.

In the scanning SET measurements, we get an additional small component of parasitic capacitance between the SET and the back-gate. This results from the fact that our SET scans at a finite height (hundreds on *nm*'s) above the MATBG. This parasitic capacitance adds a background to the measured inverse compressibility of the order of $d\mu/dV < 10^{-4}$. In the estimation of the entropy this gets doubly integrated yielding a term that depends quadratically on $\nu$. We remove this term by assuming that the entropy at $\nu = 0$ is also zero (in addition to assuming it is zero at $\nu = 4$ as discussed above). As seen in the global entropy measurements (Fig. 3b and S1b), the entropy curve that correspond to the temperature range $T = 4\text{K} - 16\text{K}$ (blue) shows that the entropy at $\nu = 0$ is smaller than $0.1 k_B$. Since local entropy measurements are performed only in this temperature range, the assumption that $s = 0$ at $\nu = 0$ is justified.

To determine the uncertainty in the local measurements of the entropy (Fig. 2e in the main text), we first extract the noise level in our measured $d\mu/dn$. We then add to our measured compressibility signal randomly distributed noise with the experimental noise variance and see how it changes the resulting entropy curve. Repeating this over a statistically significant instances of random noise gives us the error bars in our determined entropy, which are shown in Fig. S2, for the traces taken at different parallel magnetic fields (as in Fig. 2e in the main text).

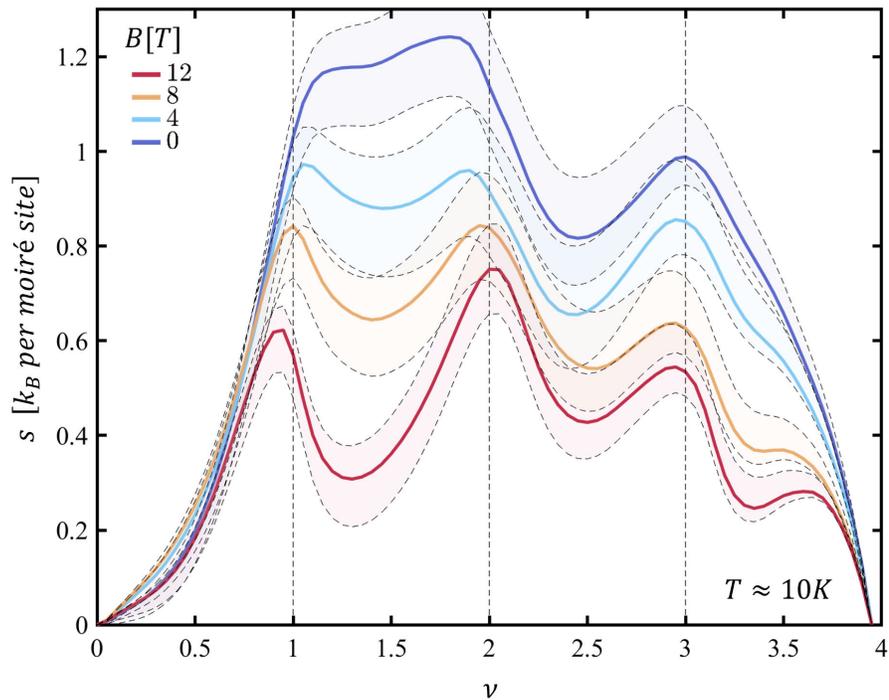

**Fig. S2: Errorbars of the measured local entropy.** Solid lines show $s$ vs. $\nu$ for serval values of $B_\parallel$. The shaded bands around each curves give the 1 sigma errorbars (see the text in this Supplementary section for details).

## SI2. Entropy of non-Interacting Dirac electrons

To get a rough understanding of the overall $\nu$ dependence of the measured entropy at high temperatures, it is useful to compare it to the entropy in a system of non-interacting Dirac bands. The curves in Fig. 3c in the main text were obtained for such a model with the a single-particle density of states that rises linearly from zero at the charge neutrality point up to the band top and bottom at $\pm W/2$, where $W$ is the bandwidth. The density of states $\rho(\varepsilon)$ for each spin/valley flavor is given by:

$$\rho(\varepsilon) = \frac{8|\varepsilon|}{W^2} \Theta\left(\frac{W}{2} - |\varepsilon|\right), \tag{1}$$

where $\Theta(x)$ is the Heaviside step function. The entropy per unit cell is then given by:

$$s(\nu, T) = -g_f k_B \int_{-\infty}^{\infty} d\varepsilon \rho(\varepsilon)\{n_F(\varepsilon)\ln[n_F(\varepsilon)] + [1 - n_F(\varepsilon)]\ln[1 - n_F(\varepsilon)]\}. \tag{2}$$

Here, $g_f = 4$ is the number of spin/valley flavors, $n_F(\varepsilon) = 1/(1 + e^{(\varepsilon-\mu)/T})$ is the Fermi-Dirac distribution, and the chemical potential is determined by solving the equation for the filling factor $\nu$, given by:

$$\nu = g_f \left[\int_{-\infty}^{\infty} d\varepsilon \rho(\varepsilon) n_F(\varepsilon) - 1\right]. \tag{3}$$

Solving Eq. (3) for $\mu(\nu, T)$ and inserting the result into (2) gives $s(\nu, T)$ shown in Fig. 3c of the main text.

## SI3. Entropy in mean-field Dirac revival model

In Refs.[1,2], we have used a simple mean-field model to describe the Dirac revival features in the compressibility. At zero temperature, this model features a cascade of phase transitions upon increasing the electron density, where the spin and valley symmetries are successively broken. At each transition, electrons of one flavor become more populated than the others. The minority flavors' densities reset to the vicinity of the charge neutrality point. This causes a sharp drop in the density of states at the Fermi level, reviving the Dirac-like density dependence of the inverse compressibility near each integer filling factor. Hence, we termed this phenomenon "Dirac revival transitions".

Here, we present a calculation of the entropy as a function of density and in-plane magnetic field within the same mean-field model. The model consists of four flavors of electrons (two valleys and two spins), each with a single-particle density of states $\rho(\varepsilon)$. The interaction, of strength $U$, is assumed to be local in real space and featureless in flavor space. The Hamiltonian is written as

$$H = \sum_{\mathbf{k},\alpha,n} (\varepsilon_{\alpha n \mathbf{k}} - \mu)\psi^\dagger_{\alpha n \mathbf{k}}\psi_{\alpha n \mathbf{k}} + H_{\text{int}}, \tag{4}$$

where $\alpha = \{K\uparrow, K\downarrow, K'\uparrow, K'\downarrow\}$ is a spin/flavor index, $n = 1,2$ labels the conduction and valence bands, $\varepsilon_{\alpha m \mathbf{k}}$ are the band dispersions (that are valley and $n$ dependent but spin independent), and the interaction Hamiltonian is given by:

$$H_{\text{int}} = \frac{U}{2N}\sum_{\alpha\neq\beta}\sum_{\{n_i\},\{\mathbf{k}_i\},\mathbf{G}} \delta_{\mathbf{k}_1+\mathbf{k}_2-\mathbf{k}_3-\mathbf{k}_4+\mathbf{G}}\, \psi^\dagger_{\alpha n_1 \mathbf{k}_1}\psi^\dagger_{\beta n_2 \mathbf{k}_2}\psi_{\beta n_3 \mathbf{k}_3}\psi_{\alpha n_4 \mathbf{k}_4}. \tag{5}$$

Here, $N$ is the number of unit cells, and $\mathbf{G}$ is a reciprocal lattice vector. The interaction couples only electrons of different spin/valley flavors, since it is assumed to be delta function-like in real space. Then, by the Pauli principle, two electrons of the same spin and valley cannot occupy the same point in real space, and do not interact. This captures the exchange part of the interaction, which favors spin or valley polarization. Including an intra-flavor term $J$, as in Ref[2]., does not change the results for the entropy shown below.

We analyze the system within a Hartree-Fock mean-field approximation, allowing for an arbitrary filling of each flavor, but no other form of broken symmetry. We use a mean-field Hamiltonian of the form:

$$H_{\text{MF}} = \sum_{\mathbf{k},\alpha,n}(\varepsilon_{\alpha n \mathbf{k}} - \mu - \mu_\alpha)\psi^\dagger_{\alpha n \mathbf{k}}\psi_{\alpha n \mathbf{k}}, \tag{6}$$

with variational parameters $\mu_\alpha$, and minimize the grand potential of the trial density matrix $\hat{\rho} = \frac{e^{-H_{MF}/T}}{\text{Tr}[e^{-H_{MF}/T}]}$. The variational grand potential per unit cell is given by

$$\Omega_{\text{MF}} = \sum_\alpha f(\mu_\alpha + \mu) + \frac{U}{2}\sum_{\alpha \neq \beta} v(\mu_\alpha + \mu)v(\mu_\beta + \mu) + \sum_\alpha \mu_\alpha v(\mu_\alpha + \mu) \quad (7)$$

where

$$f(\mu) = -T \int_{-\infty}^{\infty} d\varepsilon\, \rho(\varepsilon) \left[\log\left(1 + e^{-\frac{\varepsilon-\mu}{T}}\right) + \frac{\varepsilon-\mu}{T}\Theta(-\varepsilon)\right],$$

$$v(\mu) = \int_{-\infty}^{\infty} d\varepsilon\, \rho(\varepsilon) \left(\frac{1}{1+e^{(\varepsilon-\mu)/T}} - \Theta(-\varepsilon)\right).$$

Here, $\rho(\varepsilon) = \frac{1}{N}\sum_{\mathbf{k}} \delta(\varepsilon - \varepsilon_{\mathbf{k}})$ is the density of states of each flavor. Minimizing Eq. (6) with respect to $\mu_\alpha$, we obtain a variational estimate for $\Omega(\mu, T)$. The entropy can then be obtained though $s = -\frac{\partial \Omega}{\partial T}$. Following Ref[1], we use a simple linear model for the density of states, given in Eq. (1). Using different models for the density of states does not alter the results qualitatively.

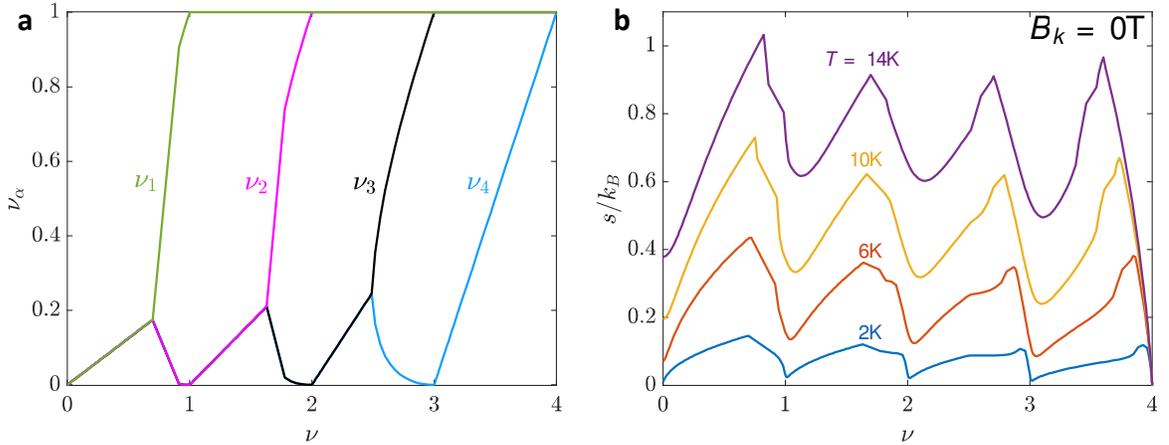

**Figure S3: Mean-field calculation. a.** Partial occupations $v_\alpha$ of each valley/spin flavor as a function of total filling factor $v$, at $T = 0$, $B_\parallel = 0$, showing a cascade of flavor symmetry breaking transitions near each integer filling. **b.** Entropy as a function of $v$ at $B_\parallel = 0$, for different temperatures. The dips in the entropy correspond to the resetting of some of the spin/valley flavors back to the charge neutrality point ($v_\alpha = 0$), while others are fully filled. At these points the density of states at the Fermi level is minimal.

We expect that at low temperatures, this approximation, built on a density matrix corresponding to a non-interacting Hamiltonian with self-consistently determined $\mu_\alpha$'s, will exhibit an entropy that is essentially $s = \frac{\pi^2}{3}\sum_\alpha \rho(\mu + \mu_\alpha)T$. Hence, the entropy is proportional to the total density of states at the Fermi level.

Fig. S3a shows the partial filling factors of each flavor as a function of the total filling factor at zero temperature, choosing $W = 2U = 300$K. The results do not change qualitatively for different values of $U/W$, as long as $2U$ and $W$ are comparable[1,2]. As seen in the figure, near charge neutrality, all four flavors start filling equally as the density is raised. Before $\nu = 1$ is reached, a phase transition occurs, in which one flavor suddenly becomes more populated than the others. When the majority flavor reaches $\nu_\alpha = 1$, the other flavors are reset to the vicinity of the charge neutrality point, and then begin filling again equally as the density is raised, until another phase transition is encountered. This is the cascade of revivals described in Refs [1,2].

In Fig. S3b, we present the entropy per unit cell $s(\nu, T)$ computed from the same model, as a function of $\nu$ for different temperatures. Thus, the entropy show clearly the revival transitions, visible as sharp dips in the entropy near each integer filling. The dips are explained by the fact that the total density of states at the Fermi level is minimal at these fillings. This $\nu$ dependence of the entropy resembles the one measured at a high field, $B_\parallel = 12$T (Fig. 2e), suggesting that the mean-field description captures the essential part of the physics there. On the other hand, the entropy measured at $B_\parallel = 0$T (fig. 2e) is quantitatively different than the one obtained here, emphasizing the imporant role of fluctuating free moments which are not included in the mean-field model.

We note that the partial fillings as a function of $\nu$ at the elevated temperatures are not strongly modified compared to those at $T = 0$, shown in Fig. S3a, although the positions of the phase transitions shift slightly with temperature.

## SI4. The effect of a magnetic field on the entropy in a mean-field model without free spins

A Zeeman field can be included in the Hamiltonian (4) by adding the following term:

$$H_Z = -E_Z \sum_{\mathbf{k},\alpha,n} \sigma_\alpha \psi^\dagger_{\alpha n \mathbf{k}} \psi_{\alpha n \mathbf{k}}, \qquad (8)$$

where $E_Z = \mu_B B_\parallel$ is the Zeeman energy, and $\sigma_\alpha$ is the spin projection of electrons of flavor $\alpha$ along the magnetic field. To account for the Zeeman field in the mean-field calculation, we replace $\mu \to \mu + E_Z \sigma_\alpha$ in Eqs. (6) and (7).

The entropy vs. $\nu$ at $T = 10$K in the presence of different in-plane magnetic fields is shown in Fig. S4. As seen in the figure, the effect of a field of up to $B_\parallel = 12$T is quite small, decreasing the entropy by at most $0.1 k_B$ relative to the $B_\parallel = 0$ value near the maxima of the entropy before the integer fillings. The change in the entropy away from the maxima due to the field is even smaller.

Comparing the mean-field results to the experimentally measured entropy (Fig. 2e in the main text), we see that the calculated entropy is in rough qualitative agreement with the experimental one at $B_\parallel = 12$T and $T \approx 10$K, showing a similar peak structure near each integer filling. The overall magnitude of the calculated entropy at $B_\parallel = 12$T is also similar to the measured one. However, the calculated entropy at $B_\parallel = 0$ is very different from the measured entropy. In particular, unlike in the calculation, the measured entropy does not drop after $\nu = 1$, but rather remains nearly constant at a high value. Moreover, the measured entropy is strongly field dependent for $\nu > 1$, whereas the calculated one is weakly field dependent at all $\nu$. We ascribe this failure of the mean-field model to the appearance of nearly-free magnetic moments (as discussed in detail in the main text). These free moments, that onset near $\nu = 1$, fluctuate strongly at low magnetic fields, an effect which is not captured in mean-field theory. Upon applying a strong Zeeman field, these fluctuations are quenched (as seen experimentally by the dramatic decrease in the entropy), and mean-field theory may be adequate.

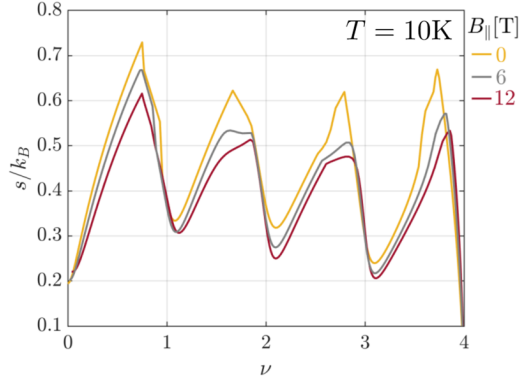

**Figure S4: Effect of an in-plane magnetic field on the entropy within the mean-field model.** In this calculation, the temperature is $T = 10\text{K}$. The different curves are for $B_\parallel = 0\text{T}, 6\text{T}, 12\text{T}$. The entropy depends only weakly on field, in contrast to the experiment. As explained above, the mean-field approximation does not capture the strong magnetic fluctuations present in the experiment at $\nu > 1$.

## SI5. Tracking $\nu_R$ using different features of the $d\mu/dn$ jump

In the main text, the transition from high to low compressibility near $\nu = 1$ was tracked by following the midpoint of the rise in $d\mu/dn$. Since the rise is fastest around its midpoint, this procedure gives us excellent resolution in defining the filling factor that corresponds to this rise, of about $\delta\nu_R \sim 0.005$. We note, however, that the overall width of the rise in filling factor can be significantly larger, and in extreme cases can even reach $\Delta\nu \approx 0.2$. It is thus necessary to check whether tracking different features of the transition as a function of magnetic field or temperature will lead to similar conclusions.

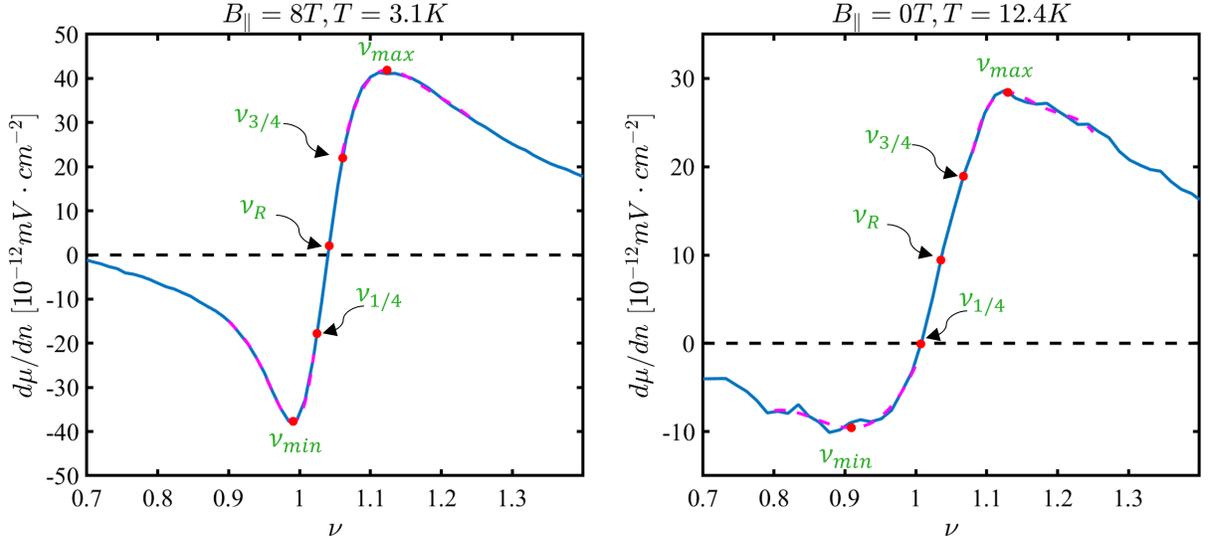

**Fig. S5: $d\mu/dn$ rise at for different $T$, $B_\parallel$. a.** Measured rise in $d\mu/dn$ near $\nu = 1$ at $B_\parallel = 8T$ and $T = 3.1K$. **b.** Same for $B_\parallel = 0T$ and $T = 12.4K$. The filling factors that correspond to the minimum and maximum of the rise, $\nu_{min}$ and $\nu_{max}$ are identified using a fit to a 4$^{th}$ order polynomial around the relevant regions (dashed purple). Also labeled are the filling factors at the midpoint of the rise, $\nu_R$, at quarter of the rise, $\nu_{1/4}$, and at three quarters of the rise, $\nu_{3/4}$.

In Fig. S5 we show two examples: the first (panel a), measured at $B_\parallel = 8T$ and $T = 3.1K$, shows a rather sharp rise. In the second (panel b), measured at $B_\parallel = 0T$ and $T = 12.4K$, the rise is more gradual. In general, similar to what is shown in these two representative measurements, we see that lower fields or higher temperatures smear the $d\mu/dn$ rise. To check how sensitive are the results shown in Fig. 4 of the main text to the choice of the definition of the location on the rise in $d\mu/dn$, we repeat the analysis with different criteria for the chosen location. Since $\nu_{min}$ and $\nu_{max}$ have large uncertainties, especially at high temperatures and low fields, we follow instead the filling factors at one quarter of the rise, $\nu_{1/4}$, and three quarters of the rise, $\nu_{3/4}$. The uncertainties in determining the latter are still low enough to make significant observations, and their tracking can still identify whether the observed features are tied to a specific part of the rise. Fig. S6 shows the extracted $\nu_R, \nu_{1/4}, \nu_{3/4}$ and $\nu_{d\mu/dn=0}$, plotted as a function of $T$ at $B_\parallel = 0T$ and $B_\parallel = 12T$. This figure should be compared with Fig. 4d in the main text.

While there are quantitative difference between the curves obtained by the different methods, we can see that in the overall dependence and the essential features in all the curves agree. For example, we see that at $B_\parallel = 12T$, independently of the method used, $\nu_R$ increases with temperature at low temperatures, reaching a maximum, and then starts decreasing with increasing temperature at high temperatures, where the crossover occurs at $T \approx 9K$.

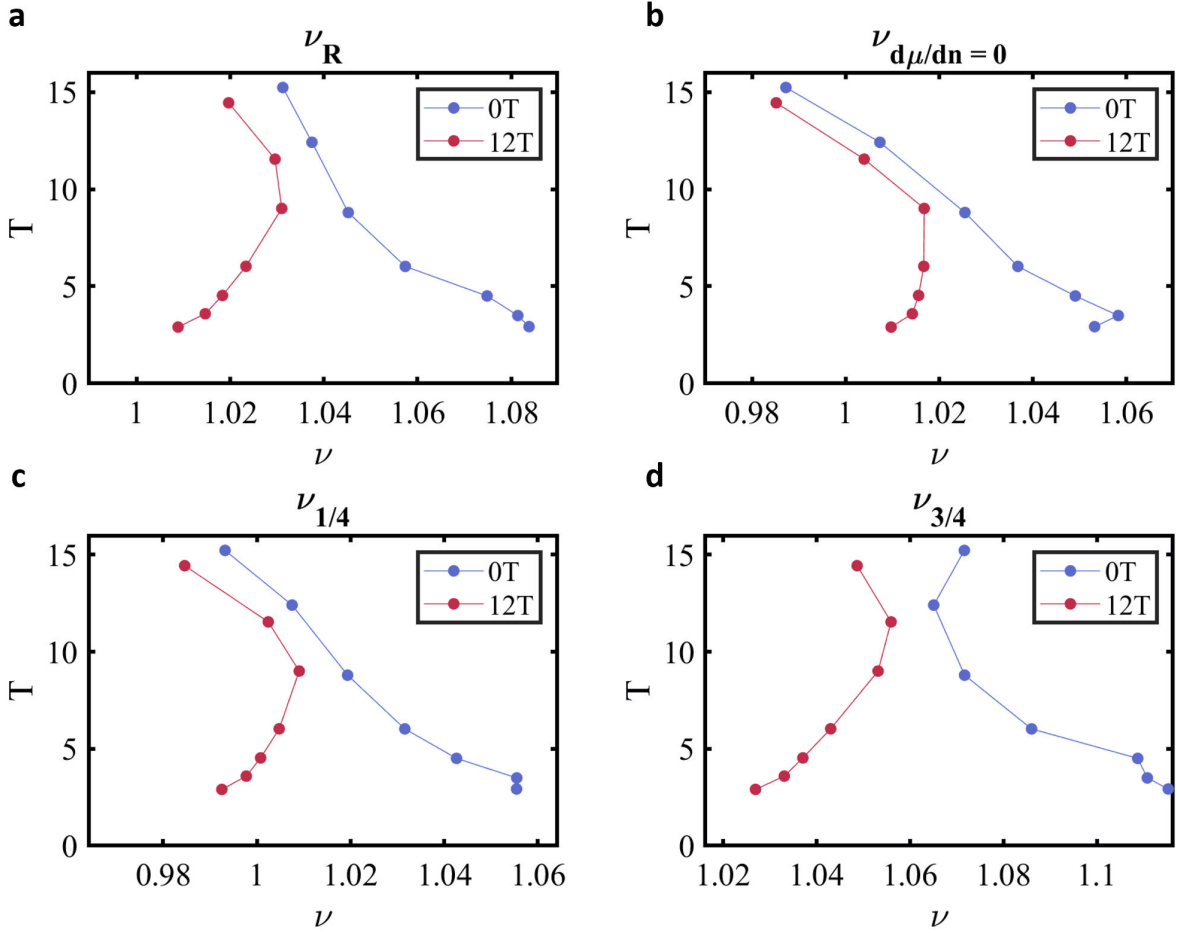

**Fig. S6: Tracking different features of the $d\mu/dn$ rise near $\nu = 1$.** a. $\nu_R$, b. $\nu_{d\mu/dn=0}$, c. $\nu_{1/4}$, d. $\nu_{3/4}$ (as defined in Fig. S5 and in this section's text) as a function of $B_\parallel$ and $T$.

## SI6. Thermodynamic model for Fermi liquid to free moment phase transition

Here, we describe the simple thermodynamic model we used in the main text to describe the first order phase transition.

The experiment is done under conditions where the temperature $T$, parallel magnetic field $B_\parallel$, and gate voltage $v_g$ are fixed. The appropriate thermodynamic potential to be minimized under these conditions is the grand canonical potential, $\Omega(v_g, T, B_\parallel)$. It is convenient to express the gate voltage in terms of the equivalent filling factor, $\nu_0 = \frac{1}{e} c_g v_g$ ($c_g$ is the geometric capacitance from the MATBG to the gate per moiré unit cell). For clarity, it is useful to derive the grand canonical potential starting from the free energy $f$, which is a function of the filling factor $\nu$, and then obtain $\Omega$ by a Legendre transformation.

Our simple model postulates the existence of a first order transition between two phases. The first phase is a relatively simple metallic phase, which we model as a Fermi liquid. The second phase is characterized by the existence of free moments. This phase is also metallic, although its density of states is lower than that of the first phase. We assume that in the second phase, there is one free spin per unit cell, coexisting with metallic Fermi liquid electrons.

The free energies per moiré unit cell of the two phases are chosen as follows:

$$f_i(\nu, T, B_\parallel) = \varepsilon_i + \frac{1}{2}\left(\frac{e^2}{c_g} + \frac{1}{\kappa_i}\right)\nu^2 - \mu_i \nu - \frac{\gamma_i T^2}{2} - \frac{\chi_i B_\parallel^2}{2} - \alpha_i T \ln\left[2\cosh\left(\frac{\mu_B B_\parallel}{T}\right)\right].$$

Here, $i = 1,2$ labels the two phases, $\varepsilon_i$ and $\mu_i$ are reference energies and chemical potentials, $\kappa_i = \left(\frac{dn}{d\mu}\right)_i$ are the intrinsic compressibilities (or quantum capacitances), $\gamma_{1,2}$ are the specific heat coefficients, $\chi_i$ are the Pauli contributions to magnetic susceptibility of the itinerant electrons, and $\alpha_i$ are the concentrations of free spins per unit cell, taken to be $\alpha_1 = 0$ and $\alpha_2 = 1$ (the results do not depend sensitively on the value of $\alpha_2$, as long as it is of order unity). We have assumed that the free spins have a gyromagnetic ratio $g = 2$.

We now carry out a Legendre transformation, $\Omega = f - ev_g \nu$, minimize $\Omega$ with respect to $\nu$, and thus eliminate $\nu$ in favor of $\nu_0 = \frac{1}{e} c_g v_g$. Since in our experiment $e^2/c_g$ is much larger than $1/\kappa_i$, we keep only terms to lowest order in $\frac{c_g}{e^2 \kappa_i}$. The grand potentials of the two phases per unit cell are:

$$\Omega_i(\nu_0, T, B_\parallel) = \tilde{\varepsilon}_i - \frac{1}{2}\frac{e^2}{c_g}\nu_0^2 - \mu_i \nu_0 - \frac{\gamma_i T^2}{2} - \frac{\chi_i B_\parallel^2}{2} - \alpha_i T \ln\left[2\cosh\left(\frac{\mu_B B_\parallel}{T}\right)\right].$$

Here, $\tilde{\varepsilon}_i = \varepsilon_i - \frac{c_g}{2e^2}\mu_i^2$. In terms of $\Omega(\nu_0, T, B_\parallel)$, the thermodynamic variables are given by:

$$v = -\frac{c_g}{e^2}\frac{\partial\Omega}{\partial v_0}, \quad s = -\frac{\partial\Omega}{\partial T}, \quad m = -\frac{\partial\Omega}{\partial B_\parallel},$$

where $s$ and $m$ are the entropy and in-plane magnetization per unit cell, respectively.

The first order transition surface in the $(v_0, T, B_\parallel)$ parameter space is given by the condition $\Delta\Omega(v_0, T, B_\parallel) = \Omega_2 - \Omega_1 = 0$. The theoretical curves shown in Fig. 4 of the main text were obtained using the following parameters: $\tilde{\varepsilon}_2 - \tilde{\varepsilon}_1 = 72K$, $\mu_2 - \mu_1 = 64K$, and $\gamma_2 - \gamma_1 = -0.0331 K^{-1}$. The negative sign of $\gamma_2 - \gamma_1$ corresponds to the fact that the density of states of itinerant carriers in the free moment phase is lower than that of the simple metallic phase. For simplicity, we neglect the Pauli contribution $\chi_i$ to the magnetic susceptibility, which is negligible compared to the free moment contribution.

Under these assumptions, the surface of the first order transition can be simply expressed as:

$$v_0^* = \frac{1}{\mu_2 - \mu_1}\left\{\tilde{\varepsilon}_2 - \tilde{\varepsilon}_1 - \frac{1}{2}(\gamma_2 - \gamma_1)T^{*2} - T^*\ln\left[2\cosh\left(\frac{\mu_B B_\parallel^*}{T^*}\right)\right]\right\},$$

where $v_0^*$, $T^*$, and $B_\parallel^*$ denote the equivalent filling factor, temperature, and magnetic field of a point on the transition surface.

The Clausius-Clapeyron relations along the transition surface can be obtained by differentiating $\Delta\Omega$:

$$d\Delta\Omega = -\frac{e^2}{2c_g}\Delta v\, dv_0^* - \Delta s\, dT^* - \Delta m\, dB_\parallel^*.$$

Here, $\Delta v = v_2 - v_1$, $\Delta s = s_2 - s_1$, and $\Delta m = m_2 - m_1$ are the jumps in the filling factor, entropy, and magnetization across the transition, respectively. Along a $v_0^* = const.$ contour of the transition surface, get the relation

$$\left(\frac{\partial T^*}{\partial B_\parallel^*}\right)_{v_0^*} = -\frac{\Delta m}{\Delta s},$$

which is the relation we used in SI7, with $v_0^*$, $T^*$, and $B_\parallel^*$ identified as the filling factor ($v_R$), temperature, and magnetic field at the Dirac revival point.

## SI7. Anti-correlation between entropy and magnetization.

The jump in compressibility seen at $v_R$ is sharp, but not discontinuous, as one may naively expect from a first order phase transition. Indeed, in the presence of long-range Coulomb interactions and disorder in two dimensions, a first order transition is not expected to be sharp. If we assume that the revival transition at $v = 1$ represent a smeared first-order phase transition, we can derive from the shape of the phase boundary the relation between magnetization and entropy. We demonstrated this relation by analyzing the slope of the phase boundary via the Clausius-Clapeyron equation: $\Delta m/\Delta s = -(\partial T/\partial B_\parallel)_{v_R}$. Here, $\Delta s$ and $\Delta m$ are the differences in the entropy and magnetization per moiré unit cell between the free moment and the Fermi liquid phases, and $(\partial T/\partial B_\parallel)_{v_R}$ is the derivative of the transition temperature with respect to magnetic field at constant $v_R$. To obtain the ratio $\Delta m/\Delta s$ we reconstruct such equi-$v_R$ contours by fitting a polynomial surface in the $B_\parallel$ and $T$ plane to the measured points, and extract the slope of the contour lines at different points (Fig. S7). Consider point A in Fig. S7: At this point, $(\partial T/\partial B_\parallel)_{v_R} \approx 0$. The Clausius-Clapeyron equations then imply that $\Delta m \approx 0$. In contrast, at point B, the equal $v_R$ contours are nearly vertical, implying that $\Delta s \approx 0$. This clear anti-correlation between $\Delta s$ and $\Delta m$ follows naturally from our simple model, where both $\Delta s$ and $\Delta m$ originate from the same free moments, that are either strongly thermally fluctuating, or polarized along the magnetic field. At point C, the contour has a positive slope, from which we deduce that $\Delta s < 0, \Delta m > 0$.

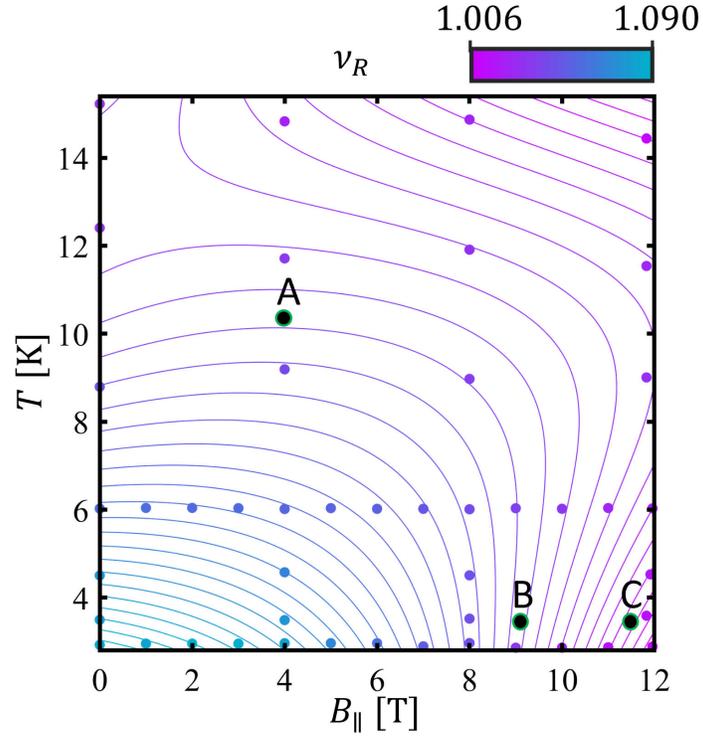

**Fig. S7: Anti-correlation between the entropy and magnetization as determined from the boundary surface curvature.** Measured $\nu_R$ as a function of $B_\parallel$ and $T$ (colored dots). The contours are obtained from a fit of these dots to a polynomial surface (3rd order in $T$ and 2nd order in $B_\parallel$). The slope of the contours in this ($B_\parallel, T$) plane gives via the Clausius-Clapeyron relation the ratio of the magnetization and entropy jumps across the transition, $\Delta m/\Delta s = -(\partial T/\partial B_\parallel)_{\nu_R}$. Visibly, in the point labeled A the contours are horizontal, implying $\Delta m \approx 0$. At point B the contours are vertical and thus $\Delta s \approx 0$. The crossover occurs along a diagonal line that correspond to the polarization of the free moments. At point C, the contour has a positive slope, from which we deduce that $\Delta s < 0$, $\Delta m > 0$.

## SI8. Comparison of transport measurements and compressibility.

Using the multilayer device shown in Fig. 3a, we can simultaneously obtain the transport resistances and the chemical potential of MATBG. Fig. S8a shows the longitudinal resistance $R_{xx}$ versus $\nu$ at different temperatures from 1K to 70K. The peaks in resistance near $\nu = -1$ denoted by the blue dots start appearing at a finite temperature of $\sim 5$K, and subsequently move to lower absolute value of filling factor as the temperature increases. The Hall coefficient and density, as shown in Fig. S8b and c, also show a similar trend. The shift of the resistive peak at $\nu = -1$ has been attributed[3] to a Pomeranchuk-like mechanism, similar to the one near $\nu = 1$.

The shift of the peak at $\nu = +1$, on the other hand, is much smaller, as was also observed in Device 1 shown in Fig. 1. Indeed, from our analysis in Fig. 4, the shift of the $\nu = +1$ state as a function of temperature is on the order of $\Delta \nu = 0.06$, which might be shadowed in the transport measurement by a moderate twist angle inhomogeneity on the order of ±0.02°.

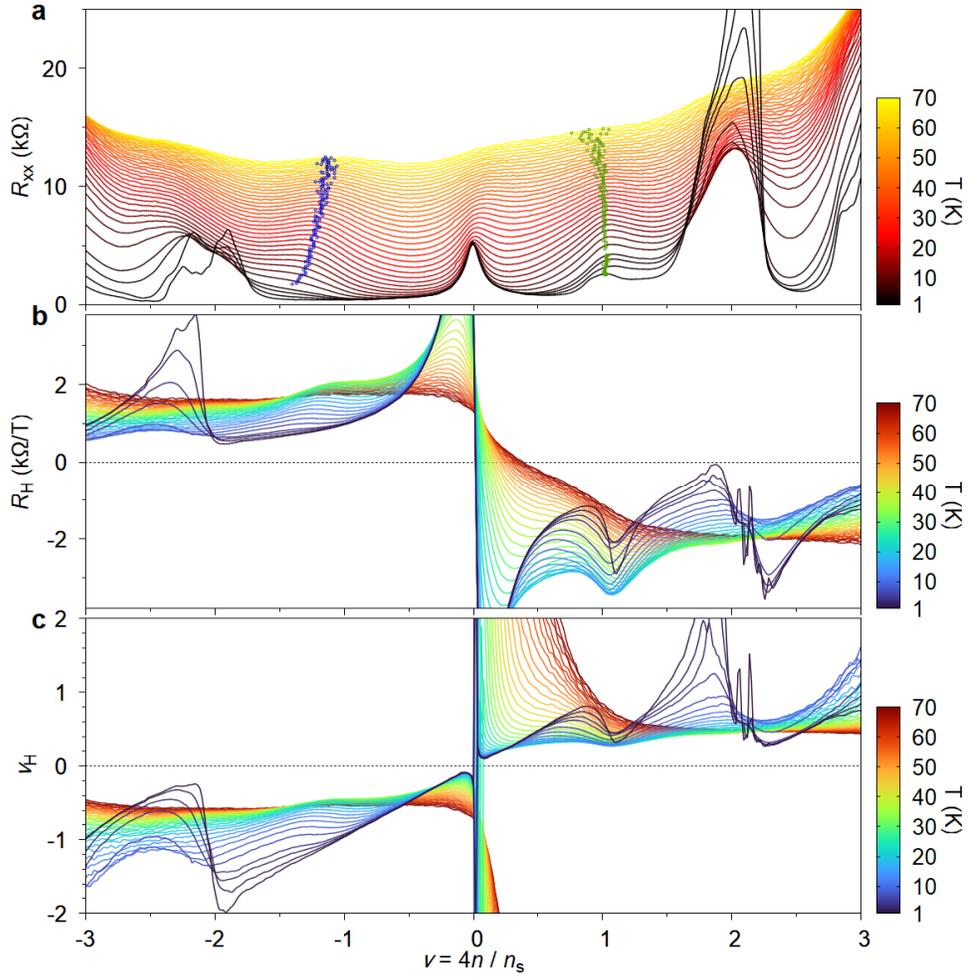

**Figure S8: Transport characterization of MATBG from $1-70K$.** (a) Longitudenal resistance $R_{xx}$ versus $\nu$. Blue and green dots mark the peaks in resistance near $\nu = \pm 1$ after a linear background is removed at each temperature. (b-c) Hall coefficient $R_H = dR_{xy}/dB$ and the corresponding Hall density $\nu_H = (-\frac{1}{R_H e})/(\frac{n_s}{4})$ in the same range of temperatures and densities.

1. Zondiner, U. *et al.* Cascade of phase transitions and Dirac revivals in magic-angle graphene. *Nature* **582**, 203–208 (2020).

2. Park, J. M., Cao, Y., Watanabe, K., Taniguchi, T. & Jarillo-Herrero, P. Flavour Hund's Coupling, Correlated Chern Gaps, and Diffusivity in Moir'e Flat Bands. *Arxiv* 2008.12296 (2020).

3. Saito, Y. *et al.* Isospin Pomeranchuk effect and the entropy of collective excitations in twisted bilayer graphene. *ArXiv* 2008.10830 (2020).